\documentclass[aps,prapplied,twocolumn,superscriptaddress,floatfix,noeprint,10pt]{revtex4-2}
\usepackage{amssymb,amsmath,amsfonts,bm}
\usepackage{graphicx}

\usepackage[usenames,dvipsnames]{color}
\newcommand{\unit}[1]{{\,\rm #1}}

\begin{document}

\title{Realtime feedback control of charge sensing for quantum dot qubits}

\author{Takashi~Nakajima}
\email[]{nakajima.physics@icloud.com}
\affiliation{Center for Emergent Matter Science, RIKEN, 2-1 Hirosawa, Wako-shi, Saitama 351-0198, Japan}

\author{Yohei~Kojima}
\affiliation{Department of Applied Physics, University of Tokyo, 7-3-1 Hongo, Bunkyo-ku, Tokyo 113-8656, Japan}

\author{Yoshihiro~Uehara}
\affiliation{Department of Physics, Tokyo University of Science, 1-3 Kagurazaka, Shinjuku-ku, Tokyo 162-8601, Japan}

\author{Akito~Noiri}
\affiliation{Center for Emergent Matter Science, RIKEN, 2-1 Hirosawa, Wako-shi, Saitama 351-0198, Japan}

\author{Kenta~Takeda}
\affiliation{Center for Emergent Matter Science, RIKEN, 2-1 Hirosawa, Wako-shi, Saitama 351-0198, Japan}

\author{Takashi~Kobayashi}
\affiliation{Center for Emergent Matter Science, RIKEN, 2-1 Hirosawa, Wako-shi, Saitama 351-0198, Japan}

\author{Seigo~Tarucha}
\email[]{tarucha@riken.jp}
\affiliation{Center for Emergent Matter Science, RIKEN, 2-1 Hirosawa, Wako-shi, Saitama 351-0198, Japan}
\affiliation{Department of Physics, Tokyo University of Science, 1-3 Kagurazaka, Shinjuku-ku, Tokyo 162-8601, Japan}

\date{\today}

\begin{abstract}
Measurement of charge configurations in few-electron quantum dots is a vital technique for spin-based quantum information processing. While fast and high-fidelity measurement is possible by using proximal quantum dot charge sensors, their operating range is limited and prone to electrical disturbances. Here we demonstrate realtime operation of a charge sensor in a feedback loop to maintain its sensitivity suitable for fast charge sensing in a Si/SiGe double quantum dot. Disturbances to the charge sensitivity, due to variation of gate voltages for operating the quantum dot and $1/f$ charge fluctuation, are compensated by a digital PID controller with the bandwidth of $\approx 100\unit{kHz}$. The rapid automated tuning of a charge sensor enables unobstructed charge stability diagram measurement facilitating realtime quantum dot tuning and submicrosecond single-shot spin readout without compromising the performance of a charge sensor in time-consuming experiments for quantum information processing.
\end{abstract}

\maketitle

Recent remarkable advances in spin qubit experiments have been facilitated by the charge sensing technique that allows measurement of charge configurations in few-electron quantum dots (QDs). A charge configuration is typically detected by measuring the conductance change of a capacitively coupled sensor transistor\cite{Elzerman:2003yt}. Measurement of spin states is also realized by using the charge sensing in conjunction with the spin-dependent electron tunneling associated with the Zeeman splitting\cite{Elzerman2004,Morello2010}, Pauli spin blockade (PSB)\cite{Petta:2005jw,Maune2012}, or quantum Hall edge states\cite{Kiyama2016}. Single-shot spin readout can be made accurate and fast enough for fault-tolerant quantum information processing by leveraging the radio frequency (RF) reflectometry\cite{Reilly2007,Keith2019} with the PSB mechanism\cite{Barthel2009,Noiri2019a,Nakajima2017,Harvey-collard2018}.
However, there is a trade off between the charge sensitivity and the dynamic range of charge sensors. Because the operating window of charge sensors is narrower when it is operated in a few-electron regime to enhance the sensitivity, the charge sensing technique requires subtle tuning of the sensor electrostatic potential that is easily affected by the gate bias voltages and the charge fluctuation in QD devices.
It is therefore necessary to perform dedicated calibration, leading to increased complexity in multi-qubit devices.
This problem is partially resolved by the gate-based dispersive readout that does not require a sensor transistor\cite{Colless2013,West2019,Urdampilleta2019,Zheng2019}, but it trades off the sensitivity.
A spin readout technique suitable for the scalable spin qubit architecture is therefore still lacking.

\begin{figure}[hbt!]
\includegraphics[width=0.39\textwidth]{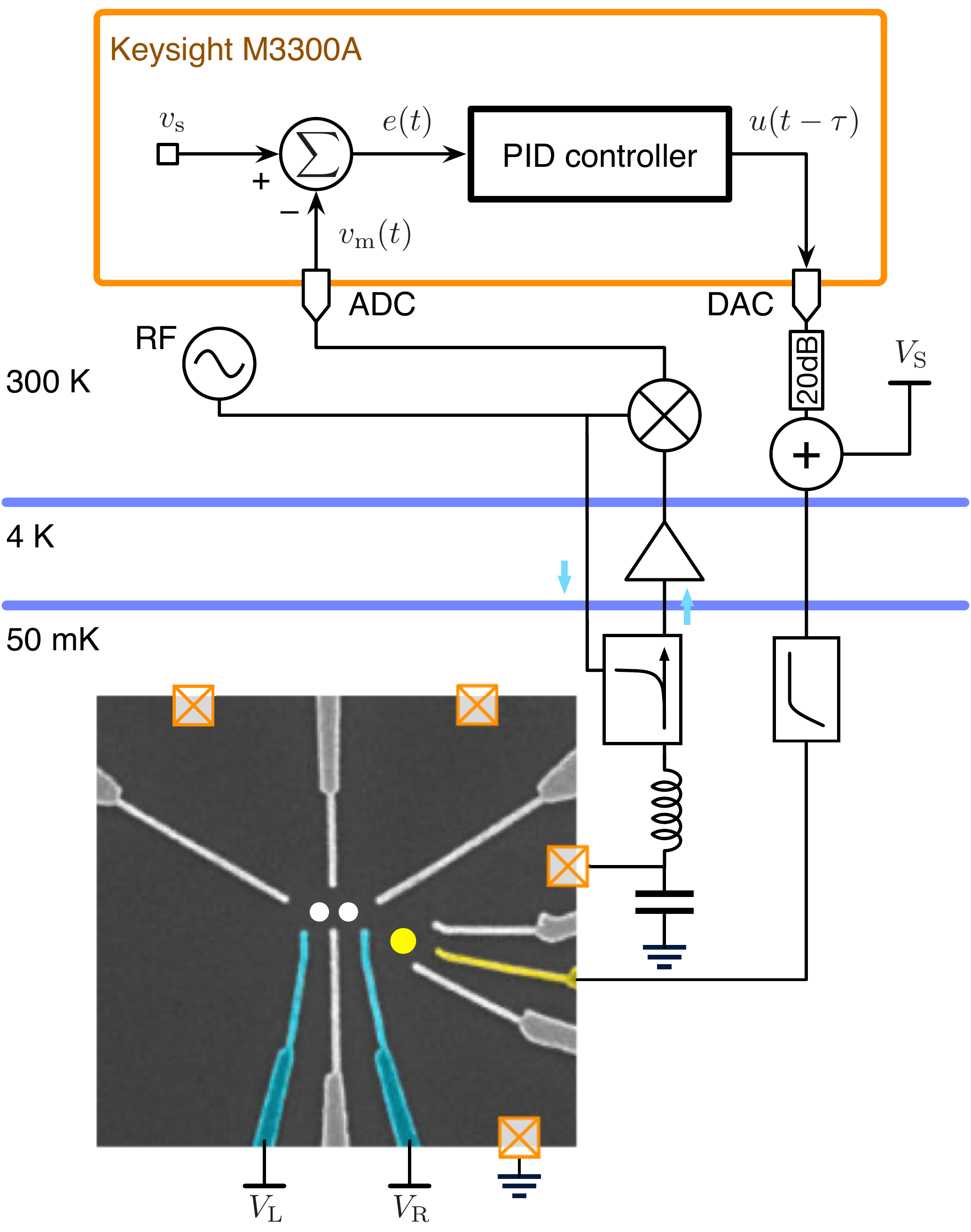}%
\caption{Schematic of the experimental setup. The Si/SiGe QD sample is mounted on the mixing chamber plate of the dilution refrigerator, where the measured electron temperature is $50\unit{mK}$ (false-colored scanning electron microscope image). The charge arrangement of the electron spin qubits confined in the DQD (white circles) is controlled by the voltages $V_\text{L,R}$ applied to the plunger gates (blue), and it is detected by the QD charge sensor (yellow circle).
The RF carrier reflected from the LC resonanct circuit is sampled by the M3300A digitizer and processed by the digital PID controller. The control voltage $u(t-\tau)$ is combined with the DC voltage $V_\text{S}$ and applied to the sensor plunger gate (yellow).
\label{fig:setup}}
\end{figure}
In this Letter, we report on the automated realtime tuning of a charge sensor for spin-qubit experiments in a Si/SiGe double quantum dot (DQD), which allows for fine control of charge sensors integrated in many-qubit devices without user intervention.
The conductance change of the charge sensor is monitored continuously and compensated by tuning the sensor plunger gate voltage\cite{Yang2011a}. This feedback loop maintains the charge sensor in sensitive conditions throughout the experiment by eliminating its unwanted variations caused by, e.g., the QD-sensor cross-capacitive coupling and the $1/f$ charge fluctuation.
By using the RF readout of the sensor conductance and a digital PID controller, we obtain a settling time of $2.2\unit{\mu s}$ allowing for compensation of slow disturbances up to $100\unit{kHz}$.
This is fast enough for live stablity diagram measurement, which significantly improves the throughput of manual tuning of QD parameters.
The automated sensor tuning in the hardware loop also allows for acquisition of unmistakable charge stability diagrams that are readily used as the input data for software-automated tuning of quantum dot arrays\cite{Mills2019a,Teske2019,Durrer2020,Lapointe-Major2020,Moon2020}.
Furthermore, we demonstrate the application of the feedback control to single-shot spin readout synchronized with qubit control pulses.
\onecolumngrid

\begin{figure*}[bth!]
\includegraphics[scale=0.75]{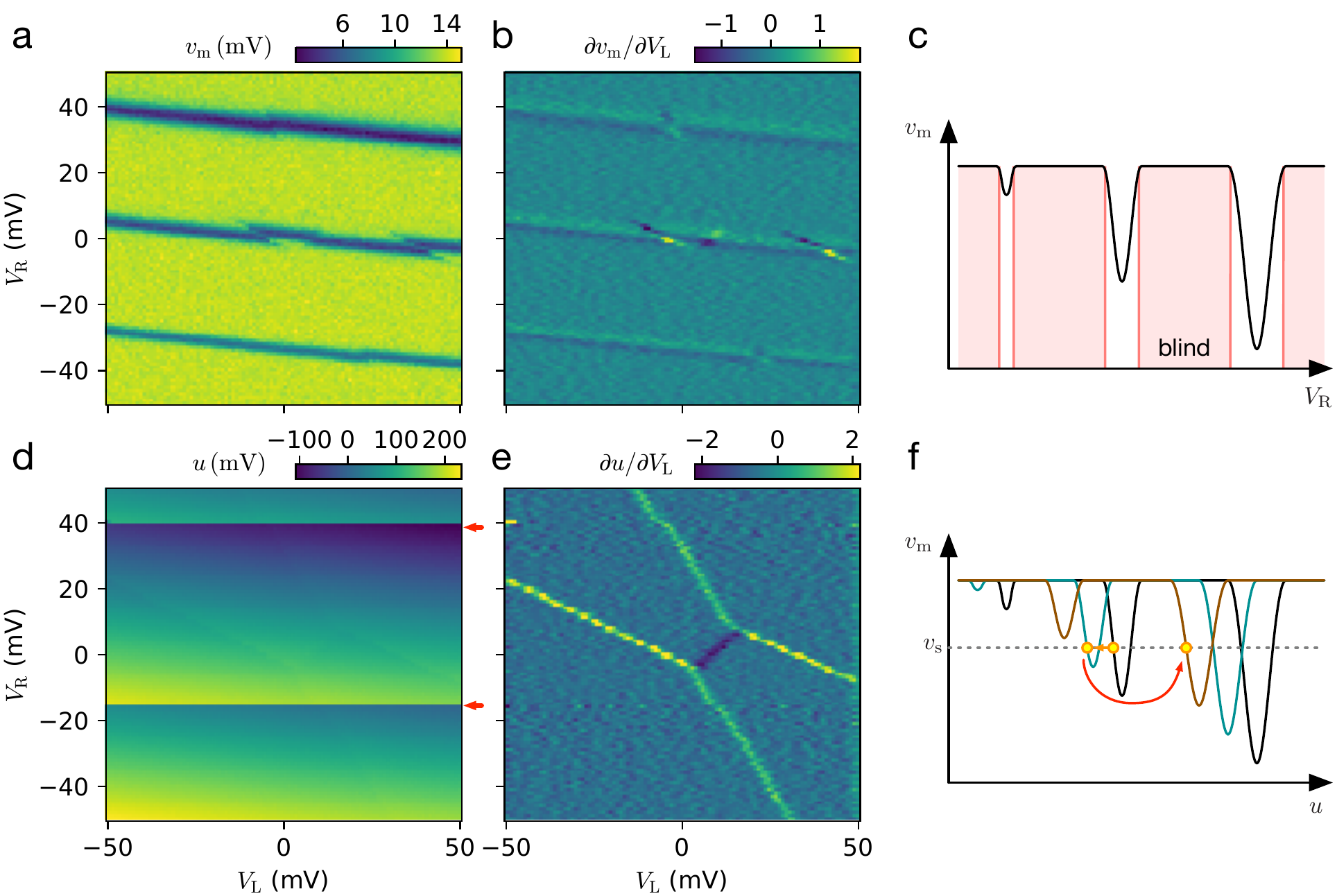}%
\caption{Stability diagrams measured with the feedback control turned off (a-c) and on (d-f). The values of $V_\text{L}$ and $V_\text{R}$ are offset from their values at the center of each diagram.
Signal is integrated for $14\unit{\mu s}$ at each data point and the whole diagram is captured at a refresh rate of about $4$ frames per second.
(a,b) Reflectometry signal $v_\text{m}$ as a function of $V_\text{L}$ and $V_\text{R}$ (a) and its derivative (b). (c) Illustration of the typical sensor signal trace during the stability diagram measurement. The sensor is sensitive to the charge arrangement only near the Coulomb peaks while it is mostly blind in the Coulomb blockade regime (red shaded region).
(d,e) Control voltage $u$ as a function of $V_\text{L}$ and $V_\text{R}$ (d) and its derivative (e), taken while the feedback control is active. The PID control is reset ($t=0$) at the lower left corner and $V_\text{L,R}$ are swept upward. Red arrows in (d) denote jumps from one Coulomb peak to another in the charge sensor. (f) Illustration of the sensor signal traces as functions of $u$ for different values of $V_\text{R}$. Yellow circles show the stable points toward which $u$ is controlled to satisfy $v_\text{m}=v_\text{s}$. As $V_\text{R}$ is increased, the signal trace shifts to the left (from black to green) and the value of $u$ at the stable point decreases. The Coulomb peak height reduces as $V_\text{R}$ is increased further (from green to brown), leading to the jumps of the stable point as indicated by the red arrows in (d).
\label{fig:stability}}
\end{figure*}
\twocolumngrid

We use a gate-defined DQD coupled with a proximal QD charge sensor fabricated on an undoped Si/SiGe quantum well wafer\cite{Takeda2016}. The charge sensing is performed by measuring the RF carrier reflected from an LC resonant circuit connected to a sensor ohmic contact\cite{Noiri2019a} (see Fig.~\ref{fig:setup}).
The demodulated RF signal $v_\text{m}$ is sampled at a rate of $100\unit{MSa/s}$ by a M3300A digitizer from Keysight Technologies. The sampled signal is fed to a tunable digital low-pass filter followed by the digital PID controller implemented in the onboard field-programmable gate array (FPGA). The control voltage $u(t)$ is given by
\begin{equation}
    u(t) = K_\text{p} e(t) + K_\text{i} \int_0^t e(s) ds + K_\text{d} \frac{de(t)}{dt},
    \label{eq:pid}
\end{equation}
where $e(t) = v_\text{s} - v_\text{m}(t)$ represents the error between $v_\text{m}$ and the desired setpoint $v_\text{s}$, while $K_\text{p}$, $K_\text{i}$, and $K_\text{d}$ are the coefficients for the proportional (P), integral (I), and derivative (D) terms. The actual output voltage is delayed by $\tau\approx 0.5\unit{us}$ due to the I/O latency of the digitizer, the digital-to-analog converter (DAC) and the FPGA logic, and updated at a rate of $100\unit{MSa/s}$. The output voltage $u(t-\tau)$ is then attenuated by $20\unit{dB}$ and added to the DC voltage $V_\text{S}$ supplied from a digital to analog converter using a home-made summing amplifier circuit. The total voltage is filtered by a Mini-Circuits VLFX-1050+ low-pass filter and applied to the plunger gate of the charge sensor that modulates the reflected signal $v_\text{m}$, thereby closing the feedback loop.

The charge stability diagrams measured with the feedback control turned on and off are shown in Fig.~\ref{fig:stability}.
When the feedback is off (Figs.~\ref{fig:stability}a-c), the electron occupation of the charge sensor is affected by $V_\text{R}$ and $V_\text{L}$ that are varied to control the charge arrangement in the DQD. The DQD charge transition lines are detectable only when the charge sensor is near the charge transitions. It is therefore often necessary to tune $V_\text{S}$ to keep the charge sensor sensitive in a desired DQD gate bias condition.
This effort of tuning $V_\text{S}$ is automated by turning on the feedback control (Figs.~\ref{fig:stability}d-f). As $V_\text{L,R}$ are varied, $u$ is controlled to keep $v_\text{m}=v_\text{s}$ where the charge sensor is sensitive. The stability diagrams in Figs.~\ref{fig:stability}d,e are obtained by monitoring the controlled output $u$, where we can clearly see the whole charge transition lines of the DQD.

\begin{figure}
\includegraphics[scale=0.75]{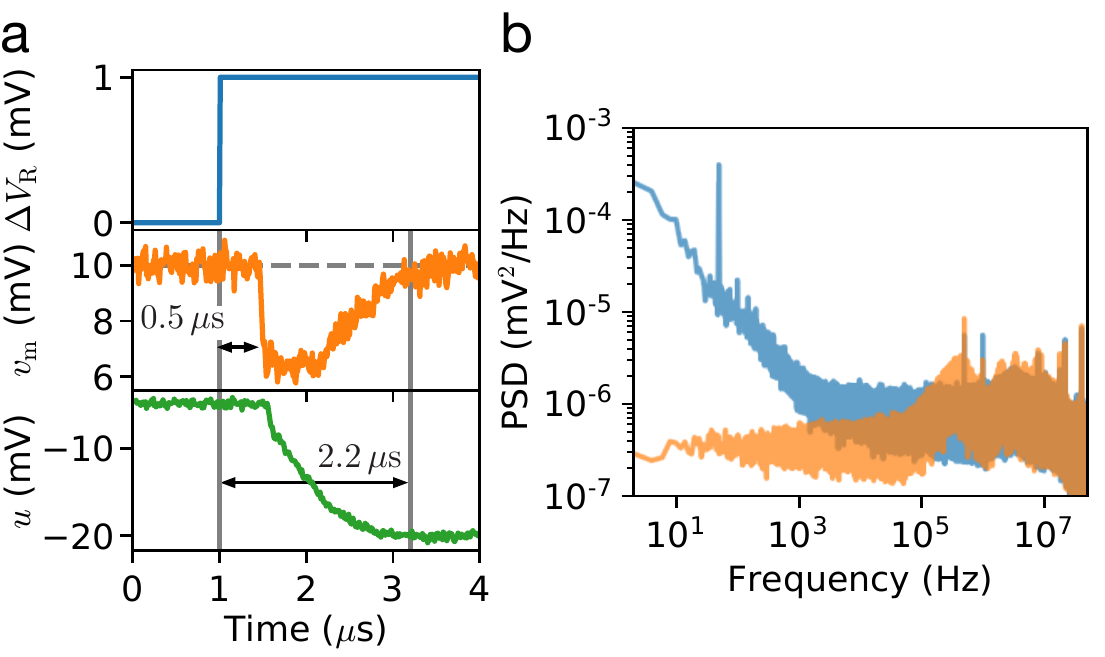}%
\caption{Performance characterization of the feedback control.
(a) Tuned-up step response of the charge sensor. A step waveform is applied to $V_\text{R}$ (upper panel) and the responses of $v_\text{m}$ (middle panel) and $u$ (lower panel) are recorded. The traces are averaged for $1,000$ trials.
(b) Noise PSDs of the charge sensor signal $v_\text{m}$. When the feedback is off (blue), we observe the typical $1/f$ tail due to the conductance fluctuation of the charge sensor along with a $50\unit{Hz}$ peak of the power line cycles. Thsese noise components are suppressed by using the feedback control (orange) up to frequencies above $100\unit{kHz}$.
\label{fig:response}}
\end{figure}

The speed and stability of the feedback control is evaluated and optimized by observing the step response of the charge sensor as shown in Fig.~\ref{fig:response}a. When a step waveform $\Delta V_\text{R}$ is added to $V_\text{R}$, we observe the response of $v_\text{m}$ after a delay of $0.5\unit{\mu s}$ mainly due to the I/O latency. Then the FPGA logic captures the error $e$ and controls $u$ to suppress the error. This effect is visible in $v_\text{m}$ after another delay of $\tau\approx 0.5\unit{\mu s}$, where $v_\text{m}$ starts to move toward $v_\text{s}$. The response of $u$ is settled and the change of $v_\text{m}$ is suppressed by $90\unit{\%}$ in $2.2\unit{\mu s}$ after the step waveform is applied, with the PID parameters $K_\text{p}=-0.80$, $K_\text{i}=-0.038$, and $K_\text{d}=0$. The response time could be further improved by decreasing $K_\text{p}$ and $K_\text{i}$, but we chose to minimize the overshoot of $v_\text{m}$ to avoid the instability caused by the nonlinearity of the charge sensor. Similarly, we chose $K_\text{d}=0$ to avoid possible instability though we did not find a significant impact by changing $K_\text{d}$ slightly.

The fast feedback control efficiently stabilizes the charge sensor by suppressing low-frequency disturbance due to the drift and $1/f$ conductance fluctuation caused by charge noise in the DQD device. As shown in Fig.~\ref{fig:response}b, the noise power spectral density (PSD) of $v_\text{m}$ is significantly suppressed by the feedback control from dc to frequencies above $100\unit{kHz}$, corresponding to the bandwidth of $\approx 1/2.2\unit{\mu s}^{-1}$. In the higher frequency range, the noise PSD only slightly increases due to the parasitic oscillation caused by the PID control. We note, however, that the suppression of the noise PSD in $v_\text{m}$ does not immediately mean the improvement of the signal-to-noise ratio (SNR) as $u$ varies to compensate the noise in $v_\text{m}$. Still, single-shot spin measurement can benefit from the stabilized charge sensing by decoupling signals from the low-frequency noise.

\begin{figure}
\includegraphics[scale=0.74]{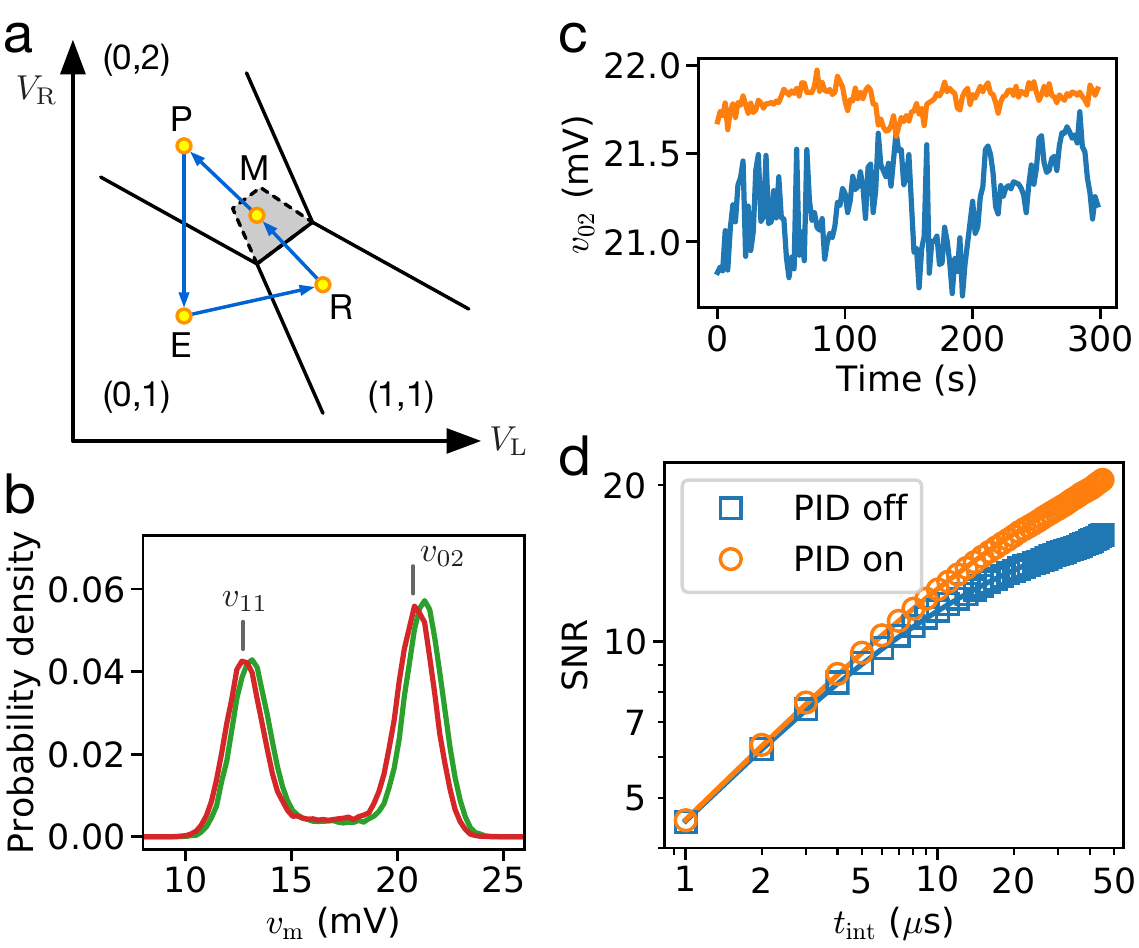}%
\caption{Single-shot spin measurement with the feedback-controlled charge sensor.
(a) 
Pulse cycle for the spin measurement using PSB\cite{Johnson:2005jw}. Solid black lines show the DQD charge transitions. The electron in the left QD is emptied (E), reloaded (R), measured (M), and parked (P) at the points marked by yellow circles. The dwell times at these points are $10\unit{\mu s}$ (E), $10\unit{\mu s}$ (R), $445\unit{\mu s}$ (M), and $35\unit{\mu s}$ (P), respectively. PSB is lifted outside the gray shaded region. The PID control is enabled or disabled at point P, while it is disabled elsewhere.
(b)
Histograms of two sets of the single-shot measurement carried out in succession without the feedback control (the PID control is disabled at point P). Each histogram is constructed from $100,000$ measurement outcomes taken by integrating $v_\text{m}$ for $t_\text{int}=5\unit{\mu s}$. The left (right) peak shows the outcomes indicating the $(1,1)$ ($(0,2)$) charge arrangement that corresponds to spin triplet (singlet). The peak positions $v_{11}$ and $v_{02}$ fluctuate with time.
(c)
Time traces of $v_{02}$ taken by turning on (orange) or off (blue) the feedback control. The values of $v_{02}$ are extracted by fitting the histograms similar to those in (b)\cite{Barthel:2010fk}.
(d)
SNRs of the single-shot measurement as functions of the integration time $t_\text{int}$ with the feedback control on (orange circles) or off (blue squares). Solid curves are the fits with $\sigma_0=1.81\pm 0.01 \unit{Vs^{-\frac{1}{2}}}$ and $t_0=0.03\pm 0.01 \unit{\mu s}$ (see the main text).
\label{fig:single-shot}}
\end{figure}

To evaluate the improvement in single-shot spin measurement, a singlet-triplet readout using the PSB effect is performed with the pulse cycle shown in Fig.~\ref{fig:single-shot}a. Figure~\ref{fig:single-shot}b shows typical histograms constructed from single-shot data taken by integrating $v_\text{m}$ for $t_\text{int}=5\unit{\mu s}$ at point M without the feedback control. The outcomes near the left (right) peak centered around $v_\text{m}=v_{11}$ ($v_{02}$) indicate the $(n_\text{L},n_\text{R})=(1,1)$ ($(0,2)$) charge arrangement that corresponds to spin triplet (singlet). Here $n_\text{L}$ ($n_\text{R}$) indicates the number of electrons in the left (right) QD. The separation of the two peaks is large enough to distinguish singlet and triplet outcomes, but the values of $v_{11}$ and $v_{02}$ fluctuate due to the first-order drift of the charge sensor. In addition, their separation $v_{02} - v_{11}$ is also affected by the second-order drift. Since the sensor drift accumulates in a long experiment, it leads to broadening of the histogram and degradation of the charge sensitivity.

In principle, one can resolve this problem by using the feedback control and measuring $u$ instead of $v_\text{m}$ as demonstrated for the stability diagram measurement in Fig.~\ref{fig:stability}. However, this approach has a few drawbacks in practice. First, the settling time of $u$ is an order of magnitude longer than the shortest single-shot measurement time achievable in a similar setup\cite{Barthel:2010fk}, though it is still fast enough for high-fidelity spin measurements\cite{Morello2010,Eng2015}. Second, application of the control pulse such as the one in Fig.~\ref{fig:single-shot}a may cause a steep response in the charge sensor and failure of the feedback loop.
To avoid these problems while taking advantage of the feedback, we interrupt the PID control synchronously with the control pulse cycle depicted in Fig.~\ref{fig:single-shot}a. In this scheme, the single-shot measurement is performed by taking $v_\text{m}$ at point M while the PID control is disabled. In the next step, the PID control is enabled at point P, where PSB is lifted and the charge sensor is stabilized for sensing the $(0,2)$ charge arrangement. Then the emptying and reloading steps follow with the PID control disabled again. Figure~\ref{fig:single-shot}c shows that the fluctuation of $v_{02}$ is successfully suppressed in this scheme.

The noise suppression improves the SNR in the single-shot measurement defined as $\text{SNR}=|v_\text{02}-v_\text{11}|/\sigma$, where $\sigma^2$ is the variance of each peak in the histogram. Figure~\ref{fig:single-shot}d shows the SNRs with the feedback turned on or off as functions of the integration time $t_\text{int}$. We fit the SNR curves to $\sigma=\sqrt{\sigma_0^2 / [t_\text{int} + t_0] + \sigma_\text{d}^2(t_\text{acq})}$, where $\sigma_0^2$ represents the white-noise broadening, $t_0$ accounts for the measurement bandwidth\cite{Barthel:2010fk}, and $\sigma_\text{d}^2(t_\text{acq})$ is the contribution from the low-frequency noise which increases with the total data acquisition time $t_\text{acq}$. For $t_\text{acq}=300\unit{s}$ used in the present experiment, we find that the SNR eventually saturates for longer $t_\text{int}$ with $\sigma_\text{d}^2=0.20\unit{mV}^2$ when the feedback is turned off. If the ideal $1/f$ noise persists in a longer experiment of $t_\text{acq}=24\unit{hours}$, $\sigma_\text{d}^2$ amounts to $0.27\unit{mV}^2$. The feedback control reduces the low-frequency noise contribution to $\sigma_\text{d}^2=0.10\unit{mV}^2$ and brings notable improvement in the SNR. This improvement, however, does not have a significant impact on the single-shot readout fidelity as long as the SNR is large enough. On the other hand, the reduction of $\sigma_\text{d}^2$ is important for a smaller SNR, which is the case when a charge sensor probes a charge arrangement in farther quantum dots or an azimuthal charge movement. As an example, consider single-shot spin measurement with $|v_{02}-v_{11}|=2\unit{mV}$ for $t_\text{int}=100\unit{\mu s}$ (shorter than the typical spin lifetime $T_1>1\unit{ms}$), which is a setup used to probe a weak charge sensing signal. In this case, the readout error is as large as $1.9\%$ for $t_\text{acq}=300\unit{s}$ and $3.5\%$ for $t_\text{acq}=24\unit{hours}$ without the feedback, while it can be improved to $0.3\%$ with the feedback. In addition to the $1/f$ noise, a charge sensor may be affected by random switching noise in a long experiment. The switching noise is also compensated by the feedback control shown in Fig.~\ref{fig:single-shot}. This feedback can be done with little overhead because the feedback settling time of $2.2\unit{\mu s}$ is shorter than typical control pulse cycles (ranging from a few $\unit{\mu s}$ to tens of $\unit{ms}$). The feedback control eliminates those possibilities of the fidelity degradation and allows for stable measurement without the need of routine calibration steps. The same feedback protocol is applicable to other single-shot measurement schemes including the one relying on the energy-selective tunneling to reservoirs\cite{Elzerman2004,Morello2010}.

In conclusion, we have developed a digital feedback control system for stable and reliable charge sensing in quantum dot devices. We have shown that a charge sensor can be maintained in a sensitive condition despite the crosstalk of gate voltages and charge fluctuation, exempting qubit operators from the labors of calibration. The settling time as short as $2.2\unit{\mu s}$ allows us to perform fast charge stability diagram measurments and single-shot spin readouts without bothered by noise slower than $100\unit{kHz}$. We expect that the feedback control of charge sensors is particularly useful in demanding computational tasks requiring a longer calculation time and a larger qubit system.


\acknowledgments
We thank RIKEN CEMS Emergent Matter Science Research Support Team for technical assistance.
Part of this work was financially supported by
JST CREST Grant Numbers JPMJCR15N2 and JPMJCR1675,
JST PRESTO Grant Number JPMJPR2017,
MEXT Quantum Leap Flagship Program (MEXT Q-LEAP) Grant Number JPMXS0118069228,
JSPS KAKENHI Grant Numbers JP18H01819 and JP19K14640,
and The Murata Science Foundation.

%


\end{document}